# Failure Tolerance of the Human Lumbar Spine in Combined Compression and Flexion Loading

Sophia K. Tushak*[a], John Paul Donlon[a], Bronislaw D. Gepner[a], Aida Chebbi[a], Bengt Pipkorn[b], Jason J. Hallman[c], Jason L. Forman[a], Jason R. Kerrigan[a]

[a]Department of Mechanical and Aerospace Engineering, University of Virginia, Charlottesville, VA, United States

[b]Autoliv Research, Sweden

[c]Toyota Motor Engineering & Manufacturing North America, United States

*Corresponding Author:

E-mail: skt5ay@virginia.edu

Tel: +1.434.297.8079

Address: 4040 Lewis and Clark Dr., Charlottesville, VA, 22911, USA

**Keywords**: lumbar spine, failure tolerance, compression and flexion, motor vehicle crashes

**Abstract:**

Vehicle safety systems have substantially decreased motor vehicle crash-related injuries and fatalities, but injuries to the lumbar spine still have been reported. Experimental and computational analyses of upright and, particularly, reclined occupants in frontal crashes have shown that the lumbar spine can be subjected to axial compression followed by combined compression-flexion loading. Lumbar spine failure tolerance in combined compression-flexion has not been widely explored in the literature. Therefore, the goal of this study was to measure the failure tolerance of the lumbar spine in combined compression and flexion. Forty 3-vertebra lumbar spine segments were pre-loaded with axial compression and then subjected to dynamic flexion bending until failure. Clinically relevant middle vertebra fractures were observed in twenty-one of the specimens, including compression and burst fractures. The remaining nineteen specimens experienced failure at the potting grip interface. Since specimen characteristics and pre-test axial load varied widely within the sample, failure forces (mean 3.4 kN, range 1.6-5.1 kN) and moments (mean 73 Nm, range 0-181 Nm) also varied widely. Tobit univariate regressions were performed to determine the relationship between censored failure tolerance and specimen sex, segment type (upper/lower), age, and cross-sectional area. Age, sex, and cross-sectional area significantly affected failure force and moment individually ($p < 0.0024$). These data can be used to develop injury prediction tools for lumbar spine fractures and further research in future safety systems.

**Introduction:**

Motor vehicle crash (MVC)-related injuries and fatalities have dropped substantially as a result of collaborative efforts to research and understand injury mechanisms using human injury surrogates as well as advances in safety systems. These human injury surrogates include anthropomorphic test devices (ATDs) (i.e. crash test dummies), computational human body models (HBMs), and post-mortem human surrogates (PMHS). Despite these research efforts and tools, some studies have reported that injuries to the lumbar spine have not been similarly reduced (Doud et al., 2015; Kaufman et al., 2013; Pintar et al., 2012; Wang et al., 2009). Of all crash directions, frontal crashes have historically elicited a higher proportion of lumbar spine injuries (Pintar et al., 2012; Shaikh et al., 2020). In particular, compression and burst fractures were observed in the thoracolumbar junction (T11-L2) and lower lumbar (L3-L5) regions and were most prevalent at L1 (Kaufman et al., 2013; Pintar et al., 2012; Rao et al., 2014; Shaikh et al., 2020). The mechanism of these injuries previously has been reported as compression and combined compression-flexion loading (Adams and Dolan, 2011; Holdsworth, 1970; Roaf, 1960; Tran et al., 1995).

This lumbar spine injury mechanism also is relevant to research for highly autonomous vehicles (HAVs), which have been predicted to introduce new reclined seating postures for vehicle occupants (Östling et al., 2019; Jorlöv et al., 2017). In particular, studies using HBMs have indicated that reclined occupants may sustain simultaneous axial compression and flexion bending in the lumbar spine (Forman et al., 2018; Gepner et al., 2019; Ji et al., 2017; Katsuhara et al., 2017; Rawska et al., 2019), where increasing recline angle amplified the lumbar axial force and flexion moment (Boyle et al., 2019; Rawska et al., 2019). Additionally, in PMHS subjected to frontal crash pulses, compression and burst fractures have been observed in occupants positioned in upright and reclined postures (Begeman et al., 1973; Richardson et al., 2020), where they described lumbar axial compression followed by torso forward rotation, resulting in the superposition of compression and flexion in the lumbar region.

Failure tolerance in the lumbar spine has been quantified at the component level for compression (Brinkmann et al., 1989; Hutton and Adams, 1982; Ochia et al., 2003; Stemper et al., 2017; Willen et al., 1984), flexion (Arregui-Dalmases et al., 2010; Belwadi and Yang, 2008; Osvalder et al., 1990; Yoganandan et al., 1988), and combined anterior-posterior shear and flexion (Belwadi and Yang, 2008; Osvalder et al., 1993). Additionally, other studies have presented data from lumbar spine axial compression and bending tests. Yoganandan et al. (1988) axially compressed thoracolumbar specimens, but load was applied quasi-statically. Duma et al. (2006) axially compressed the lumbar spine dynamically, but sample size was small (two whole lumbar spines) and resultant moments were reported rather than flexion moments. Furthermore, high shear forces were likely induced at the specimen/potting resin (grip) interface due to stiff boundary conditions, which likely contributed to the moments measured at the load cells. Belwadi and Yang (2008) applied flexion to lumbar spine segments, but load was applied at rates far lower than those observed in MVCs. These studies do not necessarily apply the same conditions observed in prior whole body PMHS and HBM studies since loading was quasi-static (Belwadi and Yang, 2008; Yoganandan et al., 1988), which is dissimilar to the high-rate loading experienced in motor vehicle crashes, and selected boundary conditions caused kinematics and kinetics dissimilar to flexion experienced during torso pitch (Duma et al., 2006).

We hypothesize that, due to the natural lordotic curvature of the lumbar spine, superposition of flexion loading with a compressed spine will reduce the maximum compression tolerance relative to pure compression loading. However, only limited information exists in the literature to describe the failure tolerance of the human lumbar spine to dynamic combined compression-flexion like that seen when occupants are in frontal crashes, particularly when highly reclined. Understanding the failure tolerance in this loading mode can enhance our injury prediction and research tools. Thus, the goal of this study was to evaluate the lumbar spine failure tolerance in combined compression and flexion.

**Methods:**

<u>Specimen Preparation</u>

Twenty-one fresh-frozen PMHS lumbar spines (T12-L5) free from pre-existing injuries, deformities, or significant ossification and degeneration were sectioned into 40 three-vertebrae segments (T12-L2, L3-L5; average: 49 years old, range: 21-74 years old) (Table 1). All tissue donation, testing, and handling procedures were approved by the University of Virginia Institutional Review Board–Human Surrogate Use (IRB-HSU) Committee.

Initial computed tomography (CT) scans of the spines were used to determine the angles between the exposed endplates and the mid-plane of the middle vertebra in the anatomical transverse plane to facilitate alignment during potting (Figure 1). *In situ* spine posture was unknown since the specimens were procured as components, so each specimen's neutral (unloaded) posture was assumed to coincide with its frozen state. Cross-sectional area (CSA) of each middle vertebra was calculated as the area within a series of approximately 40 points along the outermost edge of the superior endplate exported from the CT scans.

Segments were carefully denuded to remove surrounding muscle and adipose tissue, leaving only the intact ligamentous lumbar spine, with intact intervertebral discs, and split into upper (T12-L2) and lower (L3-L5) segments. To maximize specimen grip in potting, wood screws were inserted into the articular processes and spinous processes. The distal half of each segment's inferior vertebral endplate was fixed to a rigid pedestal at the base of a potting cup using wood screws. A hardening resin (Fast Cast #891, Goldenwest Manufacturing, Inc., Grass Valley, CA) was used to pot the specimen with special care to avoid potting resin interaction with the facet joints and intervertebral discs. Two wood screws aligned approximately with the pedicles passed through holes in the side of the potting cup and into the specimen. Then, a similar approach was completed to pot the superior vertebra. CT scans of the prepared specimen were performed before and after testing.

Instrumentation:

The middle vertebra of each specimen was instrumented with two strain gauge rosettes (Micro-Measurements® C2A-06-062WW-350) and one acoustic sensor (Physical Acoustics® Nano-30 AE Sensor) on the lateral faces of the vertebral body using cyanoacrylate glue. An array of four motion-tracking markers was fixed to the spinous process or transverse process using wood screws. Additional motion-capture markers were rigidly attached at various locations on the fixture (Figure 2), and 3D motions were recorded using an eight-camera optoelectronic stereophotogrammetric system (Vicon MX, Oxford UK).

Mechanical Test Device:

Testing was performed on a linear acceleration single intrusion cylinder modified to provide high-rate rotational motion (Figure 2) (Alshareef et al., 2018). Inferior potting cup anterior-posterior shear forces were minimized with a linear bearing rail to facilitate a pure bending moment. The test fixture applied rotations of the superior potting cup about an adjustable center of rotation that was aligned with the inferior endplate of the inferior vertebra. Six-axis load cells were placed on the outside of both potting cups. Additionally, a triaxial accelerometer was placed on the inferior potting cup, and a six-axis inertial measurement unit (3 accelerometers and 3 angular rate sensors) package (Rudd et al., 2006) on the superior potting cup. The polarities of all instrumentation followed SAE J211 (SAE, 2014).

The adjustability features were utilized to align the potted specimen assembly with the test fixture's center of rotation and align the mid-transverse plane of the middle vertebra perpendicular to the fixture's axis of compression. Axial compression was applied quasi-statically via a spring-honeycomb system (Roberts et al., 2018) and was maintained throughout dynamic flexion. To measure the flexion tolerance at varying compression levels, tests were performed with one of three pre-test compression forces: 2200 N, 3300 N, and 4500 N. Segments were randomly assigned to compression levels, with efforts to maintain approximately equal spread in sex and age within each compression level.

Specimen Testing:

Axial compression was applied quasi-statically until the honeycomb reached its crush force. Dynamic flexion to 55° at rates up to 700 °/s was applied by the test device. Load cell, accelerometer, angular rate sensor, and strain gauge data were recorded at 10 kHz. Acoustic data were recorded at 200 kHz. Multi-angle video images and 3D motion tracking data were collected at 1000 Hz, and a camcorder recorded the quasi-static axial compression.

Data Analysis:

All sensor data were debiased, and load cell, accelerometer, and angular rate sensor data were low-pass filtered using a Channel Frequency Class (CFC) 180 filter (SAE, 2014). Strain gauge data were filtered using CFC 600, and acoustic sensor data were unfiltered. Motions of the superior load cell, inferior load cell, and middle vertebra were calculated using motion capture marker trajectories, assembly schematics, and CT segmentation (Lessley et al., 2011). A joint coordinate system was defined on the middle vertebra to describe its motion (Wu et al., 2002). Once equilibrium between both load cells was verified, the inferior load cell and reconstructed 3D motion tracking data were used to calculate the kinetic state at the origin of the middle vertebra's joint coordinate system throughout the test. Inertial compensation for the accelerating load cells had negligible effects on the signal magnitudes and introduced additional noise. Maximum and minimum principal strains were calculated for both strain gauge rosettes for injury timing purposes only.

Injury Identification and Censoring:

The time and type of injury for each specimen was determined by in-depth analysis of high-speed videos, pre- and post-test CTs, and acoustic sensor, strain gauge, force, and moment data (Figure 3). All fracture classifications were verified by a board-certified radiologist. If more than one fracture occurred, only the first fracture was considered, as any subsequent fractures may have been influenced by the initial fracture. In all cases, sensor data signals corroborated fracture timing to within one millisecond.

Failure tolerance data were characterized as uncensored, left-censored, or right-censored. Specimens in which a middle vertebrae fracture occurred at an exact, known time were characterized as uncensored. Specimens in which the middle vertebrae fracture occurred during the application of axial compression were characterized as left-censored. For these tests, failure was assumed to occur in pure compression, and failure moment was zero. Specimens in which fracture occurred in the superior or inferior vertebral bodies or within the potting resin were characterized as right censored, as middle vertebrae fractures were the target, and failures at the boundaries were likely artificial.

Statistical Analysis:

Univariate Tobit regressions that permitted integration of injury threshold censoring were performed to determine the individual relationships of age, sex, segment type, and CSA to failure force and moment. Student's t-tests were performed to determine CSA variation among sexes (unpaired) and segment types (paired). Two upper segments (950, 985) were removed from the paired t-test since they did not have a corresponding lower segment. Significance value of p=0.05 was selected.

**Results:**

In 21 of the 40 specimens, the first fracture occurred in the middle vertebra, the bending moment and axial load at the time of fracture could be determined reliably in 18 cases (uncensored), whereas the tolerance data were censored in the other three cases. These fractures consisted of ten burst fractures, four major compression fractures, six minor compression fractures, and one not further specified inferior L1 fracture, as coded following AIS (AAAM, 2018) (Table 2, Figure 4). There were thirteen L1 fractures compared to eight L4 fractures. In six of the twenty-one specimens, first fracture occurred during the quasi-static application of axial compression. In three of these specimens, the force at failure was not recorded, so the force immediately after compression was substituted (left-censored). In the remaining nineteen specimens, first fracture occurred at the grip interface in the superior or inferior vertebral bodies or within the potting resin (right-censored).

Average failure force and moment were 3405 N (range: 1580 N to 5062 N) and 73 Nm (range: 0 Nm to 181 Nm) (Table 2, Figure 5) but were below average for uncensored data and above average for censored data (Table 3). CSA of the superior endplate was positively correlated (force: p=0.013; moment: p<0.001) and age was negatively correlated (force:

p=0.0067; moment: p=0.0096) with failure force and moment (Figure 6a and Figure 6b, respectively). Failure forces and moments for male specimens were significantly higher than female specimens (force: p=0.0024; moment: p<0.001) (Figure 6c). However, failure forces and moments were not significantly different between upper and lower specimens (force: p=0.14; moment: p=0.33) (Figure 6c). CSA was significantly higher in males than females (p<0.001) and in lower than upper segments (p=0.0021) (Figure 7).

**Discussion:**

The goals of this study were to quantify the failure tolerance of the lumbar spine in combined compression and flexion loading. Twenty-one of the 40 specimens sustained clinically relevant fractures, meaning that the fractures resembled those reported in field studies. The remaining 19 did not reach forces/moments high enough to elicit the targeted middle vertebrae fractures and likely represented an underestimation of mean failure forces and moments. Fractures in this second group may have been artificial due to the stiff potting resin boundary condition and stress concentrations caused by the screws used to improve potting grip. Generally, male specimens exhibited a larger number of this failure type compared to female specimens (14 vs 5). Additionally, lower segments accounted for the majority of this failure type compared to upper segments (11 vs 8). The younger specimens (<30 years old) almost exclusively displayed this failure type (7 vs 1). Although this was not the preferred failure type, the information from these tests remained valuable because it described non-injury for the middle vertebrae, as the specimens would have sustained middle vertebrae fractures had higher combined forces and moments successfully been applied. Finally, initially failure forces could not be determined if failure occurred during the application of compression, but after this failure type occurred three times, the instrumentation and load application approach was modified so that the next three that occurred could be treated as uncensored. Thus, the selection of appropriate censoring was necessary. For the statistical analysis, the Tobit model allowed for regression of the data that accounted for data censoring, whereas a typical linear regression would weight each data point equally, or as all uncensored. While it was beneficial to investigate which factors potentially influence failure tolerance individually, univariate regressions are only a first step in understanding how specimen parameters affect failure tolerance. There could be interactions between factors that were not accounted for in the univariate models. The variations in the significant factors suggest that many more specimens would be necessary to clarify these relationships.

The failure forces and moments were lower compared to previous findings of failure tolerance of whole lumbar columns loaded in dynamic axial compression (male: 5009 N and 237 Nm, female: 5911 N and 165 Nm versus average male: 3595 N and 103 Nm and average female: 3196 N and 40 Nm) (Duma et al., 2006), which could be due to the test fixture and experimental design. As opposed to constraining both ends of the specimen, the linear rail in the test fixture of this study allowed for the release of anterior-posterior translation, thereby minimizing anterior-posterior shear forces. The stiff boundary conditions of the previous study may have caused high shear forces and moments at the specimen boundaries that may have increased the failure tolerance. Additionally, loads were measured at L5, but fracture occurred at T12 in both specimens. Moments were not resolved to the fracture location, so they were affected by the test method and cannot be interpreted as failure tolerances. Likewise, failure was assumed to be aligned with the peak force and simultaneous moment. The current study showed that failure force often preceded the peak force (30/40 specimens) and sometimes by as much as 50%.

Therefore, the loads reported by Duma et al. (2006) should be considered left-censored. Additionally, compression forces were lower in the current study likely because the spring-honeycomb system dictated relatively constant force throughout the test. Duma et al. (2006) described that their whole lumbar columns were more susceptible to "first order buckling." Instead, three-vertebrae segments were chosen for this study, which have been shown to fail at similar forces as whole lumbar columns (Stemper et al., 2017) with reduced risk for buckling. In another study that tested whole thoracolumbar columns in axial compression, the resolved failure tolerance of the one specimen that failed in the lumbar region (L1) was 1730 N and 148 Nm (Yoganandan et al., 1988), which was within the range of the current data. Similar to Duma et al. (2006), Yoganandan et al. (1988) constrained both ends, but load was applied quasi-statically. Two other studies utilized a dropping mass to supply dynamic axial compression, which resulted in similar, clinically relevant, fractures as the current study, such as compression, burst, and/or endplate fractures, and average lumbar failure force in pure compression of 5.5 kN (range: 3.8-7.9 kN) for whole lumbar columns (Stemper et al., 2017) and 8 kN (range: 6-10 kN) for three-vertebrae segments (Willen et al., 1984). Greater failure forces reported in uniaxial studies compared to those reported for combined loading in this study suggests the spine acts like a curved beam that is similarly subjected to superposition of stresses. Lastly, Belwadi and Yang (2008) applied dynamic flexion to three-vertebrae segments, and failure moments (average: 140 Nm; range: 117-164 Nm) were relatively similar to those from this study (average: 73 Nm; range: 0-181 Nm) despite a substantially lower rate of flexion bending (3.5 °/s vs. ~700 °/s). Measured compression forces were 2.15 kN on average, which was on the lower end of the current data. However, loads were measured directly from the load cell and were not transformed to the middle vertebrae, so direct comparison to the current study may not be appropriate. While it is valuable to evaluate results from those previous studies relative to those in the current study, they are inherently different from the current study in the application and measurement of loads.

Rao et al. (2014) found that in the Crash Injury Research and Engineering Network (CIREN) database over half of major lumbar injuries were compression fractures, and approximately one third were burst fractures, which was opposite of the current data in which a higher number of burst fractures occurred than major compression fractures. This suggests that alternative loading mechanisms to the one used in this study could occur in the field, such as pure compression. Another field data study querying the CIREN database described that L1 sustained the highest number of overall fractures and the highest number of burst and wedge compression fractures (Pintar et al., 2012). Alternatively, L4 sustained a higher number of minor compression fractures (Pintar et al., 2012), which might be indicative of less severe endplate fractures without any marked height loss, such as the few observed in the current study. Overall, field data vertebral body fractures categorized as major were concentrated at L1, likely because the ribcage stiffens the response of the thoracic spine, causing a stress concentration at L1. Thus, L1 fractures were a primary target in this study. In both Rao et al. (2014) and Pintar et al. (2012), approximately two-thirds of fractures occurred at the thoracolumbar junction and one-third of fractures occurred in the lower lumbar.

Male specimens were more tolerant of compression and flexion loads compared to female specimens. CSA of the male specimens also was significantly larger than that of the female specimens, suggesting that increased force- and moment-tolerance is correlated with larger CSA. Further, increasing age has been shown to decrease lumbar vertebral body strength (Bouxsein et al., 2006) and was corroborated by the findings from this study. Additionally, potential differences in bone material properties between sexes and ages could influence failure tolerance but have not been investigated in this study.

The basic biomechanical data of the failure tolerance of lumbar vertebrae in combined compression-flexion loading resulting from this study can be applied to the study of injury causation and injury prevention in automotive safety, sports, and military injury biomechanics. Specifically, this study provides information to characterize the forces and moments required to generate lumbar vertebrae fractures in frontal MVCs. As a next step, the failure tolerance may be used to develop an injury risk function to describe population response relative to influential factors, such as CSA and age. Then, future work should include additional steps for the application and validation of such an injury risk function to another test surrogate, such as an ATD or HBM.

**Conclusions:**

The test protocol allowed for repeatable, consistent combined compression-flexion loading of the lumbar spine, with failures corroborated using data from multiple sensors. Clinically relevant middle vertebrae fractures were sustained in over half of the specimens and the information from the remaining specimens must be considered censored in any statistical analysis. The transformed and inertially compensated forces and moments described failure tolerance at the middle vertebra, which varied widely and were influenced by age, sex, and vertebral body CSA. These data are ideal for injury risk function development and finite element model validation, and they also can be applied to other areas within biomechanics.

**Acknowledgments:**

Thank you to the personnel at the Center for Applied Biomechanics who provided technical support for this study. Funding for this research was provided by Autoliv and the Toyota Collaborative Safety Research Center. Representatives from both of the sponsoring institutions were coauthors and aided in the study design, data interpretation, manuscript writing, and the decision to submit the manuscript for publication.

**Conflicts of Interest:**

The authors do not have any conflicts of interest to disclose.

**Tables and Figures:**

**Table 1.** Donor information.

| Subject ID | Sex | Age | Weight (kg) | Height (cm) |
|:---:|:---:|:---:|:---:|:---:|
| **752** | M | 27 | 86.2 | 182.9 |
| **864** | F | 66 | 37.6 | 165.1 |
| **938** | M | 66 | 82.5 | 167.6 |
| **939** | F | 36 | 109.8 | 157.5 |
| **942** | F | 25 | 48.5 | 152.4 |
| **946** | M | 41 | 99.8 | 165.1 |
| **948** | M | 47 | 79.4 | 177.8 |
| **949** | M | 22 | 99.8 | 185.4 |
| **950** | F | 47 | 83.4 | 162.6 |
| **953** | F | 58 | 62.6 | 160.0 |
| **964** | F | 46 | 44.5 | 165.1 |
| **965** | F | 51 | 47.6 | 157.5 |
| **966** | F | 71 | 64 | 160.0 |
| **968** | F | 56 | 61.2 | 152.4 |
| **969** | M | 74 | 108.9 | 180.3 |
| **970** | M | 49 | 86.2 | 165.1 |
| **971** | M | 45 | 61.2 | 172.7 |
| **979** | F | 59 | 68 | 167.6 |
| **983** | M | 63 | 93.9 | 162.6 |
| **985** | M | 50 | 81.6 | 185.4 |
| **986** | M | 21 | 56.7 | 177.8 |

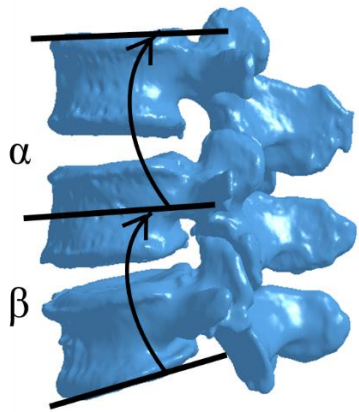

**Figure 1.** Angles calculated from initial CT scans. α is the angle between the superior endplate and the mid-plane of the middle vertebra. β is the angle between the inferior endplate and the mid-plane of the middle vertebra.

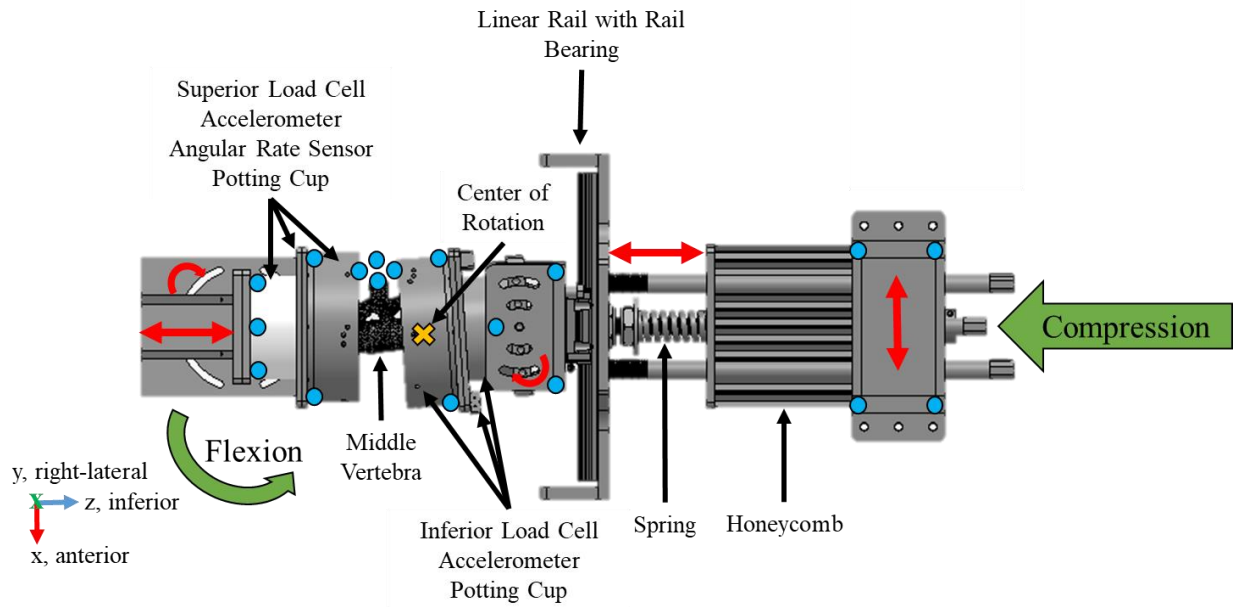

**Figure 2.** Test fixture, with rotation and translation adjustability features noted in red, Vicon markers noted in blue, and loading types noted in green.

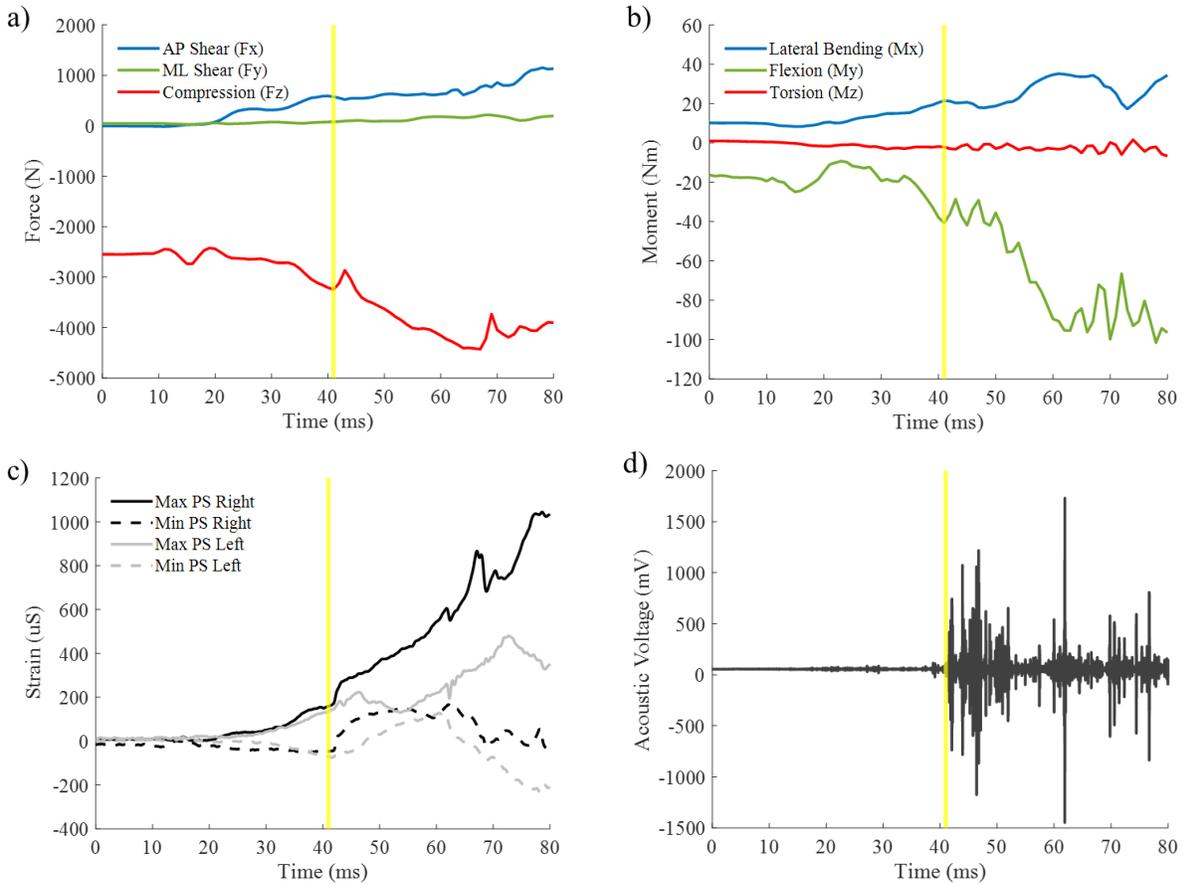

**Figure 3.** Example data for injury timing determination, including a) forces, b) moments, c) maximum and minimum principle strain (PS) for right and left strain gauges, and d) acoustic signal at the origin of the joint coordinate system located in the middle vertebrae. Timing of first fracture is noted with a yellow vertical line. Forces and moments follow the coordinate system described in Figure 2.

**Table 2.** Injury type with AIS© codes and corresponding censoring, cross-sectional area (CSA), and forces and moments at failure.

| Subject ID | Segment | Compression Level (N) | Injury Type | AIS2015 Code (AAAM, 2018) | Censoring | CSA (mm^2) | Force (N) | Moment (Nm) |
|---|---|---|---|---|---|---|---|---|
| 752 | Upper | 3300 | Superior potting failure | N/A | Right | 1049 | 3420 | 129 |
|  | Lower | 4500 | Inferior potting failure | N/A | Right | 1143 | 4668 | 165 |
| 864 | Upper | 3300 | L1 compression fx, 40% height loss, 5 mm retropulsion | 650634.3 | Uncensored | 694 | 2988 | 59 |
|  | Lower | 4500 | L4 3-column burst fx extending into bilateral facets, 20% height loss | 650636.3 | Uncensored | 834 | 3500 | 14 |
| 938 | Upper | 3300 | L1 compression flexion teardrop fx at anterior superior endplate | 650632.2 | Uncensored | 1271 | 3090 | 126 |
|  | Lower | 3300 | L4 compression fx at anterior superior endplate with wedging | 650632.2 | Uncensored | 1345 | 3241 | 153 |
| 939 | Upper | 2200 | L1 2-column burst fx, 30% height loss, 4 mm retropulsion | 650636.3 | Uncensored | 781 | 2451 | 12 |
|  | Lower | 2200 | L5 compression fx at superior endplate, no significant height loss | N/A | Right | 907 | 2444 | 42 |
| 942 | Upper | 3300 | Superior potting failure | N/A | Right | 676 | 3142 | 92 |
|  | Lower | 4500 | L4 compression fx at superior endplate, no significant height loss | 650632.2 | Uncensored | 685 | 4378 | 74 |
| 946 | Upper | 3300 | L1 3-column burst fx extending into left lamina and pedicle, 50% height loss | 650636.3 | Uncensored | 938 | 3117 | 0 |
|  | Lower | 4500 | L4 compression fx, 30% height loss | 650634.3 | Left | 932 | 3800 | 0 |
| 948 | Upper | 3300 | L1-L2 chance fx | N/A | Right | 1035 | 2963 | 76 |
|  | Lower | 3300 | Inferior potting failure | N/A | Right | 1035 | 3543 | 88 |
| 949 | Upper | 3300 | Superior potting failure | N/A | Right | 994 | 3482 | 117 |
|  | Lower | 4500 | Inferior potting failure | N/A | Right | 1331 | 5062 | 153 |
| 950 | Upper | 4500 | L1 compression fx, 50% height loss | 650634.3 | Uncensored | 824 | 3247 | 41 |
| 953 | Upper | 2200 | L1 3-column burst fx extending into the disc space and into T12 | 650636.3 | Uncensored | 915 | 1580 | 56 |
|  | Lower | 4500 | L4 3-column burst fx extending to right lamina, 25% height loss, 3 mm retropulsion | 650636.3 | Left | 961 | 4112 | 0 |
| 964 | Upper | 4500 | L1 3-column burst fx extending into bilateral lamina, 37% height loss, 7 mm retropulsion | 650636.3 | Uncensored | 835 | 2911 | 0 |
|  | Lower | 3300 | L5 compression fx at superior endplate, 11% height loss | N/A | Right | 902 | 3476 | 69 |
| 965 | Upper | 4500 | L1 3-column burst fx, >60% height loss, 10 mm retropulsion | 650636.3 | Uncensored | 839 | 3319 | 32 |
|  | Lower | 3300 | Inferior potting failure | N/A | Right | 988 | 3420 | 120 |
| 966 | Upper | 3300 | L1 compression fx with wedging, 50% height loss | 650634.3 | Left | 777 | 3145 | 0 |
|  | Lower | 3300 | L4 compression fx at inferior endplate, no significant height loss | 650632.2 | Uncensored | 875 | 3142 | 64 |

| 968 | Upper | 3300 | L2 compression fx at superior endplate, no significant height loss | N/A | Right | 738 | 3129 | 30 |
|---|---|---|---|---|---|---|---|---|
| | Lower | 3300 | L4 compression fx at superior endplate, no significant height loss | 650632.2 | Uncensored | 792 | 3076 | 29 |
| 969 | Upper | 2200 | Superior potting failure | N/A | Right | 912 | 2784 | 83 |
| | Lower | 4500 | Inferior potting failure | N/A | Right | 1041 | 4803 | 53 |
| 971 | Upper | 4500 | Superior potting failure | N/A | Right | 883 | 4312 | 105 |
| | Lower | 2200 | Inferior potting failure | N/A | Right | 1062 | 2138 | 141 |
| 972 | Upper | 4500 | L1 compression fx, <20% height loss | 650632.2 | Uncensored | 1107 | 4324 | 111 |
| | Lower | 3300 | Inferior potting failure | N/A | Right | 1056 | 3160 | 103 |
| 979 | Upper | 4500 | L1 3-column burst fx extending to bilateral facets and spinous process, 30% height loss, 2 mm retropulsion | 650636.3 | Uncensored | 880 | 3912 | 0 |
| | Lower | 4500 | L4 3-column burst fx extending to left transverse and superior articular processes, 45% height loss, 7 mm retropulsion | 650636.3 | Uncensored | 810 | 3346 | 17 |
| 983 | Upper | 2200 | L1 fx, NFS | 650698.9 | Uncensored | 998 | 2556 | 90 |
| | Lower | 2200 | Inferior potting failure | N/A | Right | 1045 | 2629 | 67 |
| 985 | Upper | 4500 | L1 3-column burst fx extending into left facet, 35% height loss, 7 mm retropulsion | 650636.3 | Uncensored | 1215 | 5042 | 55 |
| 986 | Upper | 4500 | None | N/A | Right | 1205 | 4211 | 181 |
| | Lower | 3300 | Inferior potting failure | N/A | Right | 1598 | 3142 | 167 |
| Mean (Std Dev) | | | | | | 973 (197) | 3405 (769) | 73 (54) |

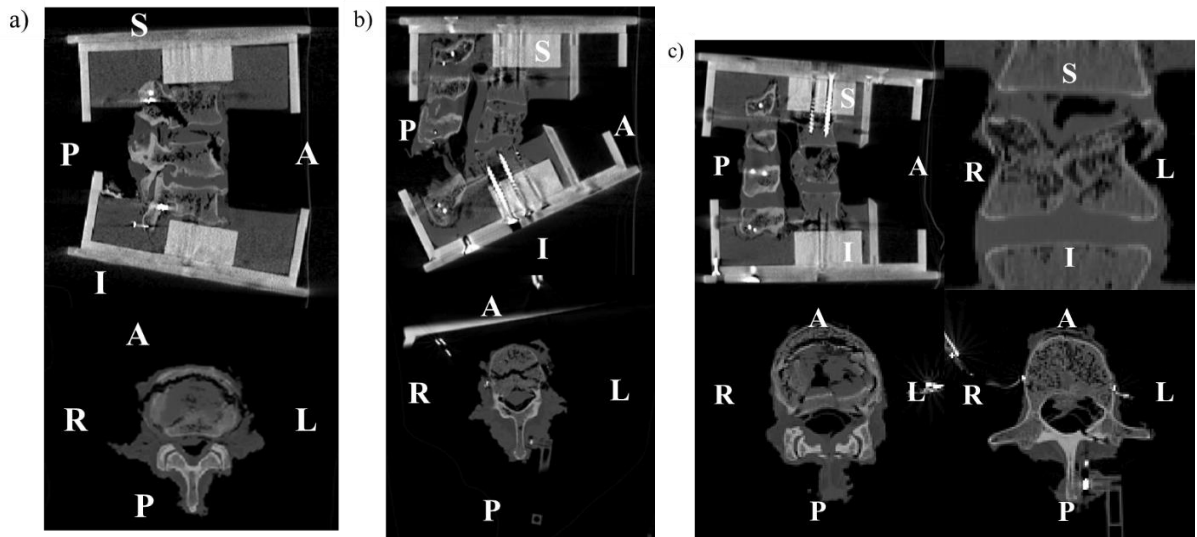

**Figure 4.** Example post-test CT images for a) minor compression, b) major compression, and c) burst fracture types, with images oriented by superior (S), inferior (I), anterior (A), posterior (P), right, (R), and left (L).

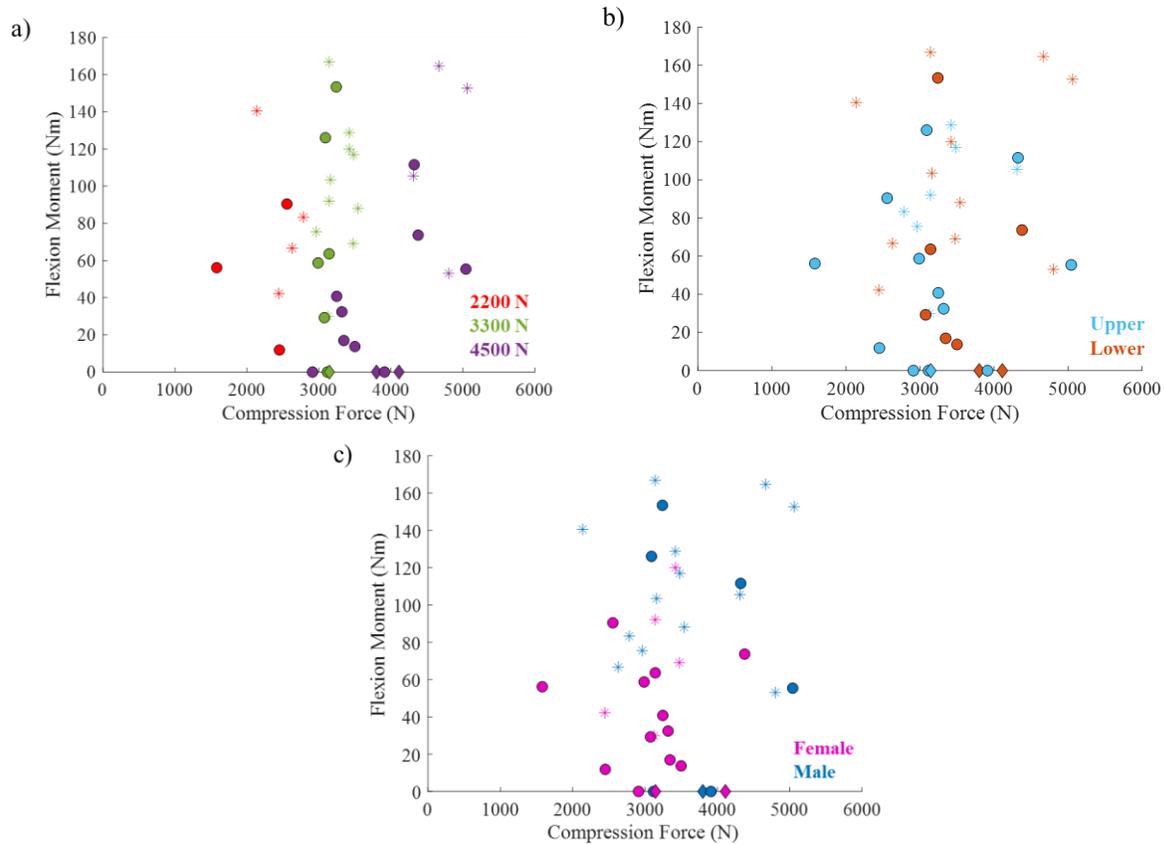

**Figure 5.** Distributions of failure forces and moments separated by a) compression level, b) segment, and c) sex, divided into uncensored (circle), left-censored (diamond), and right-censored (asterisk).

**Table 3.** Range, mean, and standard deviation for all, uncensored, left-censored, and right-censored failure forces and moments.

| | Force (N) | | | Moment (Nm) | | |
|---|---|---|---|---|---|---|
| | Range | Mean | Standard Deviation | Range | Mean | Standard Deviation |
| All | 1580 - 5062 | 3405 | 769 | 0 - 181 | 73 | 54 |
| Uncensored | 1580 - 4378 | 3290 | 778 | 0 - 126 | 52 | 45 |
| Left-Censored | 3145 - 4112 | 3686 | 494 | 0 - 0 | 0 | 0 |
| Right-Censored | 2138 - 5062 | 3470 | 806 | 30 - 181 | 104 | 44 |

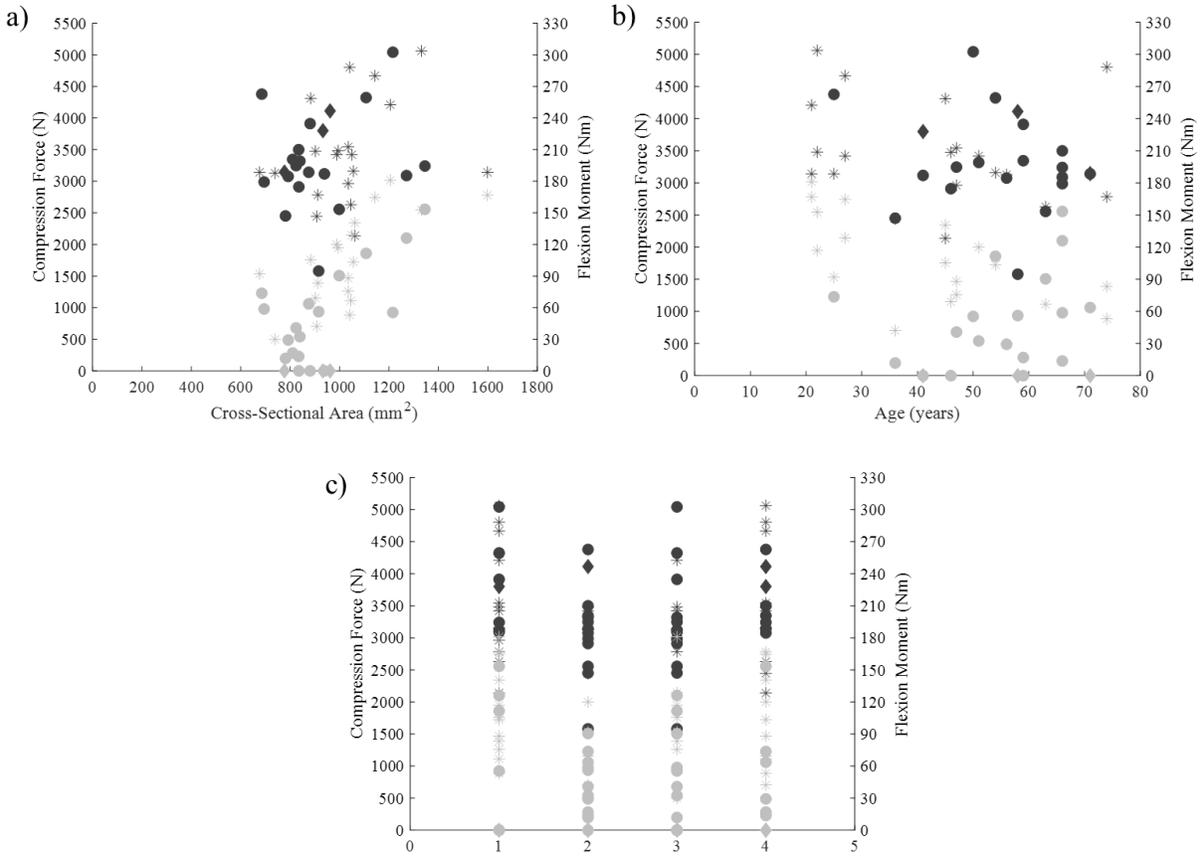

**Figure 6.** Distributions of failure forces (black) and moments (gray) versus a) CSA, b) age, and c) sex and segment type, divided into uncensored (circle), left-censored (diamond), and right-censored (asterisk).

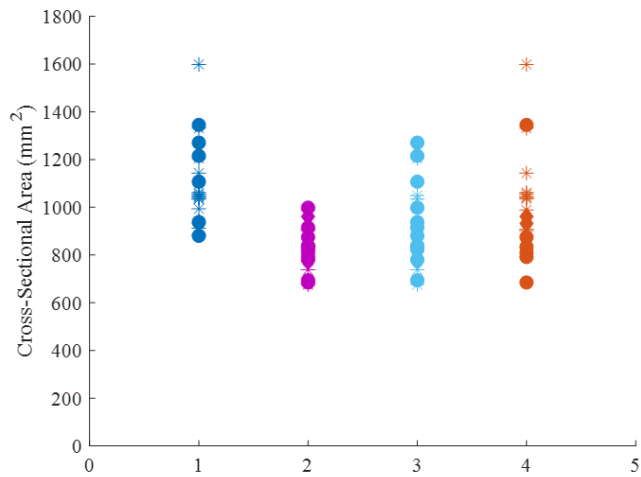

**Figure 7.** Distributions of cross-sectional areas divided into uncensored (circle), left-censored (diamond), and right-censored (asterisk).